\documentstyle[12pt]{article}

\begin{document}
\begin{titlepage}

\vspace{3cm}

\begin{centering}

{\huge Supersymmetry of the 2+1 black holes}

\vspace{2.5cm}

{\large Olivier Coussaert and Marc Henneaux$^*$\\
Facult\'e des Sciences, Universit\'e Libre de Bruxelles,\\
Campus Plaine C.P. 231, B-1050 Bruxelles, Belgium}\\

\end{centering}
\vspace{1.5cm}

\begin{abstract}
The supersymmetry properties of the asymptotically anti-de Sitter
black holes of  Einstein theory in 2+1 dimensions are investigated.
It is
shown that (i)
the zero mass black hole has two exact supersymmetries;
(ii) extreme $lM=|J|$ black holes with $M \not= 0$ have
only one; and (iii) generic black holes do not have any.
It is also argued that the zero mass hole is the ground state
of (1,1)-adS supergravity with periodic (``Ramond")
boundary conditions on the spinor fields.
\end{abstract}

\vspace{4.25cm}
{\footnotesize($^*$)Also at Centro de Estudios
Cient\'\i ficos de Santiago, Chile.}

\end{titlepage}

\vfill

\pagebreak

Among the black hole solutions of (2+1)-Einstein theory discovered
recently \cite{1}, the one with zero mass and zero angular momentum
stands apart. (i) It is the
solution with smallest mass. (ii) It has zero temperature.
(iii) It has zero entropy.
We show in this letter that it enjoys also remarkable supersymmetry
properties \cite{2}-[10].  Namely, it is the black hole solution
with the maximum
number of exact supersymmetries.  We shall first establish the
result and shall then discuss its implications \cite{rem}.

The 2+1 black hole metric is given by
\cite{1}
\begin{eqnarray}
\label{1}
\nonumber ds^2=-N^2 dt^2 + N^{-2} dt^2 + r^2 ( N^\varphi dt + d\varphi)^2\\
N^2=(r/l)^2 - M + (J/2r)^2\\
\nonumber N^\varphi=-J/2r^2
\end{eqnarray}
where $M$ and $J$ are respectively the mass and angular momentum of
the hole, and where $-l^2$ is the cosmological constant.
It can be obtained by making appropriate
identifications of the anti-de Sitter metric \cite{17}, which
corresponds to (\ref{1}) with $M=-1$ and $J=0$,
\begin{eqnarray}
\label{2}
ds^2_{adS}= -[(r/l)^2+1] dt^2 + [(r/l)^2+1]^{-1} dr^2 + r^2 d\varphi^2
\end{eqnarray}

The metric (\ref{1}) with $M \neq -1$ has
only two Killing vectors \cite{17}.
If regarded as a solution of
the equations of motion of
adS supergravity with zero gravitini,it may
possess, in addition, exact supersymmetries.  Exact supersymmetries
are by definition supersymmetry transformations leaving
the metric (\ref{1})
(with zero gravitini) invariant. The spinor parameters of these
transformations solve the ``Killing spinor equation"
\begin{eqnarray}
\label{3}
D_\lambda \psi = \frac{\epsilon}{2l} \gamma_\lambda \psi
\end{eqnarray}
where $\epsilon = 1$ or $-1$ depending on the representation
of the $\gamma$-matrices.

As it is well known,
there are two inequivalent two-dimensional irreductible
representations of
the $\gamma$-matrices in three spacetime dimensions. One
may be taken to be $\gamma^{(0)}=i
\sigma^2$, $\gamma^{(1)}=\sigma^1$ and $\gamma^{(2)}=\sigma^3$, where
the $\sigma^k$ are the Pauli matrices.
The other is given by $\gamma'^{(\lambda)}=
-\gamma^{(\lambda)}$. We shall consider here the simplest
supergravity model with negative cosmological constant involving both
representations, namely $(1,1)$ adS supergravity \cite{Ach}.

The anti-de Sitter metric (\ref{2}) possesses
 four Killing spinors, two for each inequivalent
representation of the $\gamma$-matrices. In the
radial tetrad frame
\begin{eqnarray}
\nonumber h_{(0)}=-[(r/l)^2+1]^{\frac{1}{2}} dt
\\  h_{(1)}=[(r/l)^2+1]^{-\frac{1}{2}} dr\\
\nonumber h_{(2)}=r d\varphi
\end{eqnarray}
the Killing spinors are given by
\begin{eqnarray}
\label{4}
\psi=[(\frac{N_{ads}+1}{ 2})^{\frac{1}{2}} + \epsilon
 (\frac{N_{ads}-1}{ 2})^{\frac{1}{2}}
\gamma^{(1)}] \\ \nonumber \times (cos \frac{1}{2}
(\varphi+\epsilon t/l)-sin\frac{1}{2}
(\varphi+\epsilon t/l) \gamma^{(0)}) A
\end{eqnarray}
where $A$ is a constant spinor.

Since the black hole metric can be obtained from (\ref{2})
by making appropriate
identifications, it possesses locally as many Killing spinors as anti-de
Sitter space. However, only a subset of these Killing spinors are, in
general, compatible with the identifications, i.e., invariant under
the transformations of the discrete group used in the identifications.
So, whereas all the local integrability conditions for the Killing
equations (\ref{3}) are fullfilled \cite{19}, there
 may be no Killing spinor at all
because of global reasons.

In order to discuss which Killing spinors are compatible with the
identifications, let us make a choice of coordinates in which the
identifications take a simple form. As shown in \cite{17} ,
 anti-de Sitter space
can be rewritten as
\begin{eqnarray}
\label{7}
ds^2_{ads}=-[(R/l)^2-1] dT^2 + [(R/l)^2 -1 ]^{-1} dR^2
+ R^2 d\Phi^2
\end{eqnarray}
in new coordinates $(T,R,\Phi)$ where $\Phi$ is
 not an angle but runs
over the entire real line. The Killing spinors are, in the frame
\begin{eqnarray}
\label{8}
\nonumber h_{(0)}=-[(R/l)^2-1]^{\frac{1}{2}} dT
\\ h_{(1)}=[(R/l)^2-1]^{-\frac{1}{2}} dR
\\ \nonumber h_{(2)}= R d\Phi
\end{eqnarray}
given by
\begin{eqnarray}
\label{9}
\psi=\frac{1}{\sqrt{2}}[((R/l)+1)^\frac{1}{2}+\epsilon
((R/l)-1)^\frac{1}{2}
 \gamma^{(1)}]
\\ \nonumber \times (cosh\frac{1}{2}(\Phi+\epsilon T/l)
+sinh \frac{1}{2}(\Phi+\epsilon T/l)  \gamma^{(2)})A
\end{eqnarray}
The identifications appropriate to a non extreme black hole with
angular momentum $J$ and mass $M$ ($ |J|<Ml$) have been shown
 in \cite{17} to be
\begin{eqnarray}
\label{10}
 (T,\Phi) \sim (T + J, \Phi + M), |J|<Ml
\end{eqnarray}
Since the Killing spinors are not invariant (even up to a sign) under
these identifications, they are not well defined in the quotient
space. Therefore, a generic black hole has no Killing spinor.

Let us now consider the extreme case $|J|=Ml$. The identifications
appropriate to that case are more complicated to describe in the
coordinate system where (\ref{7}) holds \cite{21}. For that reason,
 we shall
directly proceed to the explicit integration of the Killing spinor
equations in the metric (\ref{1}), where $\varphi$ is an angle.
 For definitess, we
treat the case $J=Ml$.  The case $J=-Ml$ is treated similarly.
One may take as local Lorentz frame
\begin{eqnarray}
\nonumber h_{(0)}= -N dt \\ \label{11} h_{(1)} = N^{-1} dr \\
\nonumber h_{(2)} = - \frac{Ml}{2r} dt + r d\varphi
\end{eqnarray}
One finds that the Killing spinors solutions
 of (\ref{3}) with $\epsilon = 1 $ are given by
\begin{eqnarray}
\label{12}
\psi = \frac{1}{2} [(U^{\frac{1}{2}} +
 U^{-\frac{1}{2}})+(U^{\frac{1}{2}}-U^{-\frac{1}{2}}) \gamma^{(1)}]
\\ \nonumber [1+\frac{1}{2}(\gamma^{(2)}-\gamma^{(0)})
 (\varphi + t/l)] A
\end{eqnarray}
where $A$ is a constant spinor and $U$ is given by
\begin{equation}
U=l/r ((r/l)^2-M/2).
\end{equation}
 The Killing
spinors for the other representation of the $\gamma$-matrices
are
\begin{eqnarray}
\label{14}
\psi=\frac{1}{2} (r/l)^{\frac{1}{2}}
[\alpha cosh (\sqrt{M/2} (t/l-\varphi)) +\\
\nonumber \beta sinh (\sqrt{M/2}(t/l-\varphi))]
\left( \begin{array}{c} 1 \\-1 \end{array} \right) \\
\nonumber+ \frac{1}{2}
(r/l)^{-\frac{1}{2}} \sqrt{M/2}
[\alpha sinh (\sqrt{M/2}(t/l-\varphi))
\\ \nonumber + \beta cosh (\sqrt{M/2}
 (t/l-\varphi))\left( \begin{array}{c} 1 \\ 1 \end{array} \right)
\end{eqnarray}
where  we recall that  $\gamma^{(1)}=\sigma^{(1)}$,
 and where $\alpha$ and $\beta$
are constants. The Killing spinor
(\ref{12}) is compatible with the periodicity of $\varphi$ if and
 only if the linearly growing term in $\varphi$ disappears,
i.e., if and only if
$A$ is an eigenstate
of $\gamma^{(1)}$ with eigenvalue $+1$. In that case,
(\ref{12}) does not depend on $\varphi$ and is thus manifestly
periodic.  The Killing spinor (\ref{14})
 is never periodic or
anti-periodic. There is thus only one Killing spinor for the extreme black
hole with non vanishing mass.

In the limit $M
\rightarrow 0$, one gets from each sign of $J$ a
$\varphi$-independent
Killing spinor.  These read explicitly
\begin{equation}
\label{15}
\psi_1=\frac{1}{2}(r/l)^{\frac{1}{2}}
\left( \begin{array}{c} 1 \\ 1 \end{array} \right)
\end{equation}
and
\begin{equation}
\label{16}
\psi_2=\frac{1}{2}(r/l)^{\frac{1}{2}}
\left( \begin{array}{c} 1 \\-1 \end{array}
 \right).
\end{equation}
The zero mass state has thus two exact supersymmetries.

The Killing spinors of the extreme black hole solutions have the same
asymptotic growing in $r$  as the Killing spinors of anti-de Sitter
space.  However, they are periodic in $\varphi$, while those of
anti-de Sitter space are anti-periodic.

This feature has interesting implications.
It has been established in \cite{13} that a negative cosmological
constant allows for rich asymptotics. Namely,
there exist boundary conditions on the
gravitational variables such that
the asymptotic symmetry algebra of ($2+1$)-gravity
with a negative cosmological constant is the conformal algebra in two
dimensions, i.e., twice the Virasoro algebra.  The mass and angular
momentum are respectively given by $M =l^{-1}(K_0 + L_0)$ and $J = K_0 -
L_0$, where $K_n$ and $L_n$ are the right and
left Virasoro generators.
These boundary conditions include
the black hole solutions of \cite{1}.

Now, in a spacetime with the
black hole topology $R^2 \times S^1$ \cite{17}, one
can consider spinor fields that are either periodic or
anti-periodic in $\varphi$ in the
above radial triad frames.
These different behaviours define inequivalent spinor structures
and lead to different asymptotic superalgebras for
(3+1)-supergravity with a negative cosmological constant.
The periodic case yields the Ramond graded extension of the
Virasoro algebra and will be referred to as the
``Ramond sector" for that reason. The anti-periodic case yields the
Neveu-Schwarz extension and will be called the
``Neveu-Schwarz sector".

 Just as in $3+1$ dimensions \cite{14,15,16},
the asymptotic supersymmetry algebra implies
bounds for the generators  $K_0$ and $L_0$.
The stronger ones are
\begin{equation}
\label{NS1}
K_0 = G_{1/2} G_{-1/2} + G_{-1/2} G_{1/2} - \frac {1} {2}
 \geq - \frac {1} {2},
\end{equation}
\begin{equation}
\label{NS2}
L_0 = \bar G_{1/2} \bar G_{-1/2} + \bar G_{-1/2} \bar G_{1/2}
 - \frac {1} {2}
 \geq - \frac {1} {2}
\end{equation}
for the Neveu-Schwarz case, and
\begin{equation}
\label{R1}
K_0 = F^2_0 \geq 0
\end{equation}
\begin{equation}
\label{R2}
L_0 = \bar F^2_0 \geq 0
\end{equation}
for the Ramond case. The $G_k$ and $F_k$ are the asymptotic
right supersymmetry generators,
while the $\bar G_k$ and $\bar F_k$ are
the  asymptotic left supersymmetry generators.

The exact supersymmetries of anti-de Sitter space
belong to the Neveu-Schwarz sector and are generated by the right and
left supersymmetry charges with ``frequency" $1/2$ and $-1/2$ (the
Killing spinors have that dependence on $\varphi$).
Hence, anti-de Sitter space is annihilated by $G_{1/2}$
and $\bar G_{1/2}$  and
saturates the bound for the Neveu-Schwarz case, in agreement
with $ M = - 1$.  Similarly, the zero mass hole
is invariant under the two zero-mode supersymmetries generated
by $F_0$ and $\bar F_0$ and saturates the bounds (\ref{R1})
and (\ref{R2}) of the Ramond
case, leading to $M = 0$. Accordingly, the zero mass
hole appears as the ground state of the Ramond sector.
The extreme black holes $lM=|J|$ with $M \not= 0$ saturate only one
of the bounds (\ref{R1}) or (\ref{R2}).

In the above analysis, we have set the electric charge equal to zero.
 The reason why we did not consider
charged black holes is that these appear to possess somewhat
unphysical properties in 2+1
dimensions. (i) They fail to fullfill the fall-off conditions given in
\cite{13} for asymptotically anti-de sitter spaces.
(ii) The energy $M$
is not bounded from either below or above when the charge
is different from zero.
Indeed given an arbitrarily negative mass, the solution
given in \cite{1} possess an event horizon hiding the singularity
for Q big enough. The
unboundedness of the energy renders the solutions unstable and
should imply the absence of asymptotic Killing spinors -a fortiori of
exact Killing spinors.

Once one imposes the asymptotic behaviour of \cite{13}, one must take
$Q=0$. This forces the electromagnetic field to
vanish and makes the vector
potential locally pure gauge.
The vector potential
is not necessarily
globally pure gauge, however, since the fundamental group of black
hole solutions is non trivial.
Because of the presence of non contractible loops,
the black hole spacetimes can support
non zero holonomies of $A_\lambda$, given by $A_t=0$, $A_r=0$,
$A_\varphi=Constant$. This is somewhat reminiscent of (3+1)-black holes
with axionic charge \cite{18}. It is no accident, since in 2+1
dimensions, the electromagnetic field ( and not the Kalb-Ramond field)
is dual to the axion field .

In four-dimensional Einstein-Maxwell theory with zero
 cosmological constant, the only black
holes with exact supersymmetries are the extreme Reissner-Nordstrom
black holes. These have the further remarkable property that one can
construct static, extreme multi-black hole solutions \cite{22,23},
in which the Coulomb potential exactly balance the gravitational
attraction. Both properties have sometimes been related, so that it is
natural to ask whether one can also construct static multi-black holes
solutions in 2+1 dimensions.
It turns out that this is not the case. As we shall show in detail
in a separate publication where the general static solution
of 2+1 Einstein theory
with negative cosmological constant will be constructed,
there is no static, supersymmetric,
pure multi-black hole metric without additional (undesirable)
naked branch point singularities.

To conclude, we have shown in this letter that
 the zero mass state enjoys remarkable
supersymmetry properties.  These indicate that the zero mass state is
the ground state of the Ramond sector of (1,1) adS supergravity.
Furthermore, exact supersymmetry is
associated with the precise bounds guaranteing the absence of naked
singularity (cosmic censorship), as in 3+1 dimensions \cite{11}. The
 extreme bound $lM=|J|$ yields one supersymmetry. The bound $M=0$
yields a second supersymmetry.  A detailed presentation of this work,
covering the extended $(p,q)$ adS supergravity models, will be
reported elsewhere.

\vskip 0.5 cm
{\bf Acknowledgements}

One of us (M. H.) is grateful
to Andrew Strominger
for interesting questions on the
supersymmetry properties of the charged $2+1$-black
holes and to Claudio
Teitelboim  for useful comments
and discussions. He also gratefully acknowledges the hospitality
of the Institute for Advanced Study where this
work has been partly carried out.
 O. C. is ``Chercheur I.R.S.I.A".
This research has been supported in part by research
funds from F.N.R.S. and by a research contract with the Community of
the European Communities.


\begin{thebibliography}{000}
\bibitem{2} Exact supersymmetry properties of (3+1)-black holes
have been must studied since the realization
 that extremal Reisner-Nordstrom black holes have some unbroken
supersymmetries[3-6]. These properties have been shown recently
 to play a central role in nonrenormaliztion theorems for extreme
black holes[7]. They have been generalized to dilaton gravity[8-10].
\bibitem{rem} Even though its horizon has zero length,
we shall refer to the zero mass solution as being a black hole since
the singularity, which coincides with the horizon, is
invisible to an observer at infinity.
\bibitem{19} The fulfilment of the integrability condition is quite
 obvious
in terms of the Chern-Simons formulation of (2+1)-supergravity
\cite{Ach,Witten}. Indeed, the
Killing equations then read $\nabla_\lambda \psi=0$ where $\nabla_\lambda$
 is the covariant derivative
in the $SO(2,2)$-connection. Since this connection is flat, the equations
 are integrable.
\bibitem{21} If one takes $|J|=Ml$ in the identifications (\ref{10}),
one does not get an asymptotically
anti-de Sitter black hole solution.
Nevertheless, the resulting quotient space is quite interesting: it has no
singularity, no closed time like curves,
two Killing spinors and four Killing vectors ( the vector
field used in making the identification is everywhere
spacelike and turns out to be self-dual $(J=Ml)$
or anti-self-dual $(J=-Ml)$).
\end{thebibliography}
\end{document}